\documentclass[reprint, superscriptaddress, amsmath, amssymb, aps, pra]{revtex4-1}
\usepackage{graphicx}
\usepackage{dcolumn}
\usepackage{bm}
\usepackage[dvipdfm,colorlinks,linkcolor=blue,urlcolor=blue, anchorcolor=blue,citecolor=blue]{hyperref}
\usepackage{color}
\usepackage{ulem}
\usepackage{amsmath,amsfonts,amssymb}
\usepackage{booktabs}
\usepackage{graphicx}
\usepackage{romannum}
\usepackage{amsmath}
\usepackage{amssymb}
\usepackage{mathtools}
\begin{document}
\title{Robust, fast, and efficient formation of stable tetratomic molecules from ultracold atoms via generalized stimulated Raman exact passage}
\author{Jia-Hui Zhang}
\email[]{jhzhang\_phys@nwnu.edu.cn}
\affiliation{College of Physics and Electronic Engineering Northwest Normal University, Lanzhou 730070, China}
\author{Wen-Yuan Wang}
\affiliation{College of Physics and Electronic Engineering Northwest Normal University, Lanzhou 730070, China}
\author{Fu-Quan Dou}
\affiliation{College of Physics and Electronic Engineering Northwest Normal University, Lanzhou 730070, China}

\begin{abstract}
The study of the conversion of ultracold atoms into molecules has long remained a hot topic in atomic, molecular, and optical physics. However, most prior research has focused on diatomic molecules, with relatively scarce exploration of polyatomic molecules. Here we propose a two-step strategy for the formation of stable ultracold tetratomic molecules.
We first suggest a generalized nonlinear stimulated Raman exact passage (STIREP) technique for the coherent conversion of ultracold atoms to tetratomic molecules, which is subsequently followed by a chainwise-STIREP technique to transfer the resulting molecules into a sufficiently stable ground state. Through systematic numerical analysis, we demonstrate that the proposed two-step strategy holds great potential for the robust, fast, and efficient formation of stable ultracold tetratomic molecules.
\end{abstract}
\maketitle

\section{\label{sec:level1}Introduction}
The successful creation of various ultracold diatomic molecules has stimulated growing interest in polyatomic molecules (UPMs)~\cite{Zeppenfeld2012, PhysRevLett.118.173201, doi:10.1126/science.ade6307, PhysRevLett.131.043402, Vilas2024, Chen2024}
In contrast to diatomic molecules, UPMs offer additional complexity~\cite{Carr_2009, B820045A, C002764B}, lending themselves to numerous interesting studies in, for instance, cold chemistry~\cite{Karman2024}, precision measurements~\cite{Hutzler_2020} and the realization of exotic quantum phases~\cite{PhysRevResearch.4.013235}, to name a few.
In order to tap their full potential, however, long-lived ensembles and full quantum state control serve as prerequisites~\cite{Langen2024}. Thus, the efficient conversion of ultracold atoms into stable UPMs constitutes an essential requirement, which hitherto remains challenging.

In the ongoing experiments, ultracold molecules are first prepared by association of pre-cooled atoms~\cite{RevModPhys.82.1225}, either via magneto or the photo-association (PA)~\cite{Dulieu_2009, 10.1063/1.3357286, Q2012, D1CS01040A}. Subsequently, the formed molecules should be immediately transferred to absolute ground states, it is commonly achieved via the well-known Stimulated Raman adiabatic passage (STIRAP)~\cite{PhysRevA.65.063619, PhysRevLett.101.133005} and its variants, such as multi-state STIRAP~\cite{Danzl2010, Mark2009, PhysRevA.78.021402, PhysRevA.78.033617, PhysRevA.81.013632, PhysRevA.103.033301}.
In principle, by manipulating the system to evolve along the adiabatic dark state, an efficient and robust state transfer can be obtained~\cite{RevModPhys.70.1003, RevModPhys.79.53, RevModPhys.89.015006}.
Until now, STIRAP has been demonstrated successfully~\cite{Bergmann_2019}, however, its application has been limited to the realm of diatomic molecules~\cite{PhysRevLett.98.043201, PhysRevLett.101.133005, Danzl1062, PhysRevLett.105.203001, Ospelkaus2008, PhysRevA.80.043620, PhysRevLett.122.253201, Duda2023, gjzh-8dsb}. Directly applying it to UPMs preparation remains challenging. On the other hand, researchers have developed some generalized versions of STIRAP, for instance, Feshbach/Efimov resonance-assisted STIRAP, to support theoretical investigations into UPMs preparation~\cite{PhysRevLett.99.133002, PhysRevA.79.063415, PhysRevA.83.043605, PhysRevA.85.023629}.
Nevertheless, the successful application of such techniques is limited by the adiabatic criterion, which renders the systems vulnerable to environmental factors, e.g., collisions or decay to outside states.

In order to relax limitations such as adiabaticity related to the STIRAP-based scheme,
a technique which is called shortcut-to-adiabaticity (STA)
was proposed~\cite{PhysRevLett.105.123003}.
By leveraging the STA, the adiabatic process can be expedited while ensuring that the outcomes remain consistent. One especially successful method for
 designing such fast routes is counterdiabatic driving~\cite{Demirplak2005, doi:10.1063/1.2992152, Berry_2009, pqhl-nbtk, wbbs-s8fs, PhysRevA.111.L031301}, which allows one to construct a modification of an original Hamiltonian to eliminate unwanted nonadiabatic couplings~\cite{RevModPhys.91.045001}.
Currently, STA is widely used in theory to realize fast and robust ultracold molecule formation~\cite{PhysRevA.103.023307, Zhang_2021, CAO2023113882} and control the selective vibrational population transfer of molecules~\cite{10.1063/1.4922779, Masuda2015A}.
In a recent study, Zhu $et$ $al$.~\cite{PhysRevA.103.023307} proposed a fast-forward scaling-assisted STIRAP method to study the conversion from atoms to diatomic molecules.
In another study, Vitanov developed a shortcut multi-state STIRAP technique~\cite{PhysRevA.102.023515}, which may provide an alternative ways for forming stable ground-state ultracold molecules. Very recently, a similar study has been reported~\cite{Han_2025}. Compared to the traditional three-level system, the employment of multi-level ones exhibits unique advantages; to illustrate, this strategy proves instrumental in resolving the challenge of weak Franck-Condon factors (FCFs) during the process of both atom-to-molecule conversion~\cite{PhysRevLett.89.040402, PhysRevA.81.013632} and ground-state molecular transfer~\cite{Danzl2010, PhysRevA.78.021402}.
Despite these advances, the major problem is that most prior studies usually require introducing additional couplings for eliminating non-adiabatic effects, which may pose additional experimental challenges.
On the other hand,
in 2017, Dorier $et$ $al$.~\cite{PhysRevLett.119.243902} suggested an alternative method, namely nonlinear stimulated Raman exact passage (STIREP), for the efficient atom-to-molecule conversion. This method conforms to the spirit of STA, because the control fields are determined by the specific dynamics selected to achieve exact state transfer~\cite{PhysRevA.105.032807, PhysRevA.109.062613}. This trick has been extended to the formation of triatomic molecules~\cite{CAO2024130019}. Currently, an important question in this realm is whether it is possible to extend this successful tool to form larger polyatomic molecular complexes,~\cite{PhysRevA.87.043631, PhysRevResearch.7.023187, PhysRevLett.131.043402, PhysRevLett.132.093403, PhysRevResearch.3.023163, gu2025} and generalize it to transfer of molecules into deeply-bound ground states.~\cite{PhysRevA.78.043417, Danzl2010, PhysRevA.78.021402, PhysRevA.109.013315}

The purpose of this paper is to develop a theoretical formalism that extends the existing STIREP protocols to the case that permits control of conversion from atoms to stable tetratomic molecules. We first report a generalization of the STIREP technique for controlling the conversion from ultracold atoms to tetratomic molecules.
To transfer the resulting molecules to a sufficiently stable ground state in a short time, we further develop a chainwise-STIREP (abbreviated as
C-STIREP hereafter) technique.
We present the complete explanations of the underlying mechanism and the corresponding numerical analyses. The results demonstrate that our proposed two-step strategy performs well in both the preparation of ultracold tetratomic molecules and the ground-state molecular transfer.

This paper is outlined as follows. In Sec.~\ref{sec:level2}, we formulate the model for the conversion of atoms to tetratomic molecules, followed by a discussion on the underlying mechanism of the generalized N-STIREP technique. Numerical analyses are presented. In Sec.~\ref{sec:level3}, we suggest and analysis a chainwise-STIREP for molecular transfer into stable ground state based on a five-state molecular model. Section.~\ref{sec:level4} concludes the paper and presents an outlook.

\section{\label{sec:level2}COHERENT CONVERSION FROM ATOMS TO TETRATOMIC MOLECULES VIA GENERALIZED NONLINEAR STIREP}
\subsection{\label{2.1}Model and Method}
\begin{figure}[b]
\centering{\includegraphics[width=8cm]{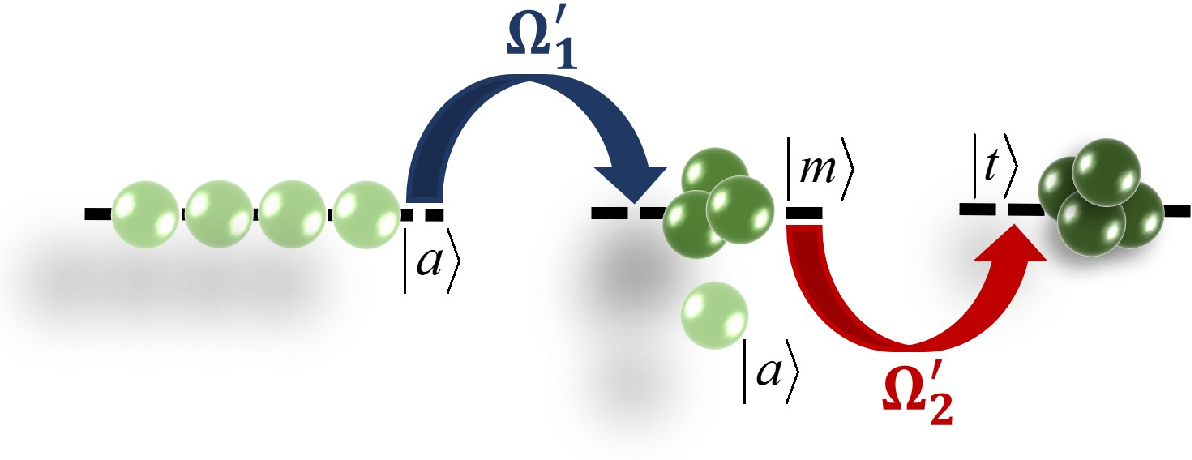}}
\caption{(Color online) Schematic showing the coherent conversion from ultracold atoms to tetratomic molecules~\cite{RevModPhys.78.483}.}
\label{fig1}
\end{figure}

As outlined in the introduction, our top priority is to prepare homonuclear tetratomic molecules.
Figure.~\ref{fig1}(a) shows a visualization of the process.
Starting from a Bose condensate of atoms, we apply the field $\Omega_{1}^{\prime}(t)$ to associate free atoms $|a\rangle$ into triatomic molecules $|m\rangle$; the other field $\Omega_{2}^{\prime}$(t), in turn, is responsible for associating pairs composed of an atom and a triatomic molecule $|m\rangle$ into tetratomic molecules $|t\rangle$. For the applied lasers, their frequencies are expressed using the single-photon and two-photon detunings $\delta(t)$
and $\Delta(t)$, respectively. We remark that the formation of triatomic molecules via the PA of free atoms has been recently reported~\cite{PhysRevA.103.022820}, whereas the formation of tetratomic molecules through the coupling of an atom-triatomic molecule pair has also been investigated~\cite{PhysRevA.77.065601}. In addition, Jones $et$ $al$.~\cite{RevModPhys.78.483} pointed out that a sufficiently dense atomic sample facilitates the combination of three atoms via PA to form triatomic molecules.

Drawing on the above theoretical advances and the well-established strategy for the stepwise coherent assembly of UPMs~\cite{PhysRevLett.99.133002}, our aim is to generalize the existing N-STIREP protocol, with the goal of enabling fast and efficient conversion of atoms to tetratomic molecules, while ensuring the occupation of the intermediate triatomic molecule remains negligible.


The Hamiltonian of this system, in the interaction picture with $\hbar=1$, takes the form of~\cite{PhysRevA.77.065601, PhysRevA.85.023629}
\begin{eqnarray}\label{1}
\begin{aligned}
\hat{\mathcal{H}}\!=
&-\int \! d\mathbf{r}\left\{\sum_{ij}\chi_{ij}^{\prime}\hat{\psi}_{i}^{\dagger}(\mathbf{r})\hat{\psi}_{j}^{\dagger}(\mathbf{r})\hat{\psi}_{j}(\mathbf{r})\hat{\psi}_{i}(\mathbf{r})\!+\!\delta\hat{\psi}_{\mathrm{m}}^{\dagger}(\mathbf{r})\hat{\psi}_{\mathrm{m}}(\mathbf{r})\right.\\
&\!+\!\Omega_{1}^{\prime}(t)\{\hat{\psi}_\mathrm{m}^\dagger(\mathbf{r})[\hat{\psi}_\mathrm{a}(\mathbf{r})]^{3}+\mathrm{H.c.}\}\!+\![\Delta(t)\!+\!\delta(t)]\hat{\psi}_\mathrm{t}^\dagger(\mathbf{r})\hat{\psi}_{\mathrm{t}}(\mathbf{r})\\
&\!-\!\Omega_{2}^{\prime}(t)[\hat{\psi}_{\mathrm{t}}^{\dagger}(\mathbf{r})\hat{\psi}_{\mathrm{m}}(\mathbf{r})\hat{\psi}_{\mathrm{a}}(\mathbf{r})+\mathrm{H.c.}]\Bigg\},
\end{aligned}
\end{eqnarray}
where $\hat{\psi}_{i}^{\dagger}(\mathbf{r})$ and $\hat{\psi}_{i}{(\mathbf{r})}$ are, respectively, the bosonic creation and annihilation operators for state $|i\rangle (i\!=\!\mathrm{a, m, t})$. In the present analysis, we focus on a uniform system, and for this reason, the usual kinetic term has been omitted~\cite{PhysRevLett.93.250403}. For systems that contain a sufficiently large number of particles, the mean-field method holds validity~\cite{PARKINS19981}. By adopting this method, we use the substitution $\hat{\psi}_{i}\rightarrow\sqrt{n}{\psi}_{i}$ ($n$ denotes the initial atomic density), and the Heisenberg motional equation can be described as
\begin{eqnarray}\label{2}
\begin{aligned}
i\dot{\psi}_\mathrm{a}\!&=\!-2K_{\mathrm{a}}(t)\psi_\mathrm{a}\!-\!3\Omega_{1}{\psi_\mathrm{m}\psi^{\ast}_\mathrm{a}}^{2}\!+\!\Omega_{2}\psi_\mathrm{t}{\psi^{*}_\mathrm{m}},\\ i\dot{\psi}_\mathrm{m}\!&=\!-2K_{\mathrm{m}}(t)\psi_\mathrm{m}\!-\!(\delta\!+\!i\gamma_\mathrm{m})\psi_\mathrm{m}\!-\!\Omega_{1}\psi_\mathrm{a}^{3}\!+\!\Omega_{2}\psi_\mathrm{t}{\psi^{\ast}_\mathrm{a}}, \\
i\dot{\psi}_\mathrm{t}\!&=\!-2K_{\mathrm{t}}(t)\psi_\mathrm{t}(t)\!-\!(\Delta\!+\!\delta\!+\!i\gamma_\mathrm{t})\psi_\mathrm{t}\!+\!\Omega_{2}\psi_\mathrm{m}\psi_\mathrm{a}.
\end{aligned}
\end{eqnarray}
In the above equations, $\chi_{ij}\!=\!n\chi_{ij}^{\prime},\Omega_{1}\!=\!n\Omega_{1}^{\prime},$ and $\Omega_{2}\!=\!\sqrt{n}\Omega_{2}^{\prime}$ denote the renormalized quantities. The decay rates $\gamma_m$ and $\gamma_t$ are introduced to phenomenologically describe the losses of the corresponding states caused by factors such as inelastic collisions and laser-induced dissociation, 
under the assumption that these decays dominate all other loss mechanisms.
The time-dependent third-order nonlinearity term $K_i(t)\!=\!\sum_j\chi_{ij}\left|\psi_j\right|^2$ characterizes elastic collisions between particles, and gives rise to the dynamical instabilities~\cite{PhysRevLett.99.223903, PhysRevA.88.063622}.

In principle, The effects of third-order nonlinearity can be eliminated by resonance locked inverse engineering~\cite{PhysRevLett.119.243902, PhysRevA.108.042610}. To this end, a general parametrization with the time-dependent phase $\alpha$ is introduced, i.e.,
\begin{eqnarray}\label{3}
\begin{aligned}
\psi_{\mathrm{a}}\!&\mapsto\!\psi_{\mathrm{a}}e^{i\alpha(t)}, \\ \psi_{\mathrm{m}}\!&\mapsto\!\psi_{\mathrm{m}}e^{\mathrm{3}i\alpha(t)}, \\ \psi_{\mathrm{t}}\!&\mapsto\!\psi_{\mathrm{t}}e^{\mathrm{4}i\alpha(t)}.
\end{aligned}
\end{eqnarray}
In this way, the above equation can be expressed as
\begin{eqnarray}\label{4}
\begin{aligned}
i\dot{\psi}_\mathrm{a}\!&=\!-[2K_\mathrm{a}(t)\!-\!\dot{\alpha}]\psi_\mathrm{a}\!-\!3\Omega_{1}{\psi_\mathrm{m}\psi^{\ast}_\mathrm{a}}^{2}\!+\!\Omega_{2}\psi_\mathrm{t}{\psi^{*}_\mathrm{m}},\\ i\dot{\psi}_\mathrm{m}\!&=\!-\left[2K_\mathrm{m}(t)\!-\!3\dot{\alpha}\!+\!\delta\right]\psi_\mathrm{m}\!-\!i\gamma_\mathrm{m}\psi_\mathrm{m}\!-\!\Omega_{1}\psi_\mathrm{a}^{3}\!+\!\Omega_{2}\psi_\mathrm{t}{\psi^{\ast}_\mathrm{a}}, \\
i\dot{\psi}_\mathrm{t}\!&=\!-[2K_\mathrm{t}(t)\!-\!4\dot{\alpha}\!+\!\Delta\!+\!\delta]\psi_\mathrm{t}(t)\!-\!i\gamma_\mathrm{t}\psi_\mathrm{t}\!+\!\Omega_{2}\psi_\mathrm{m}\psi_\mathrm{a}.
\end{aligned}
\end{eqnarray}
With the substitution $\dot{\alpha}=2K_\mathrm{a}(t)$ implemented, the condition for resonance locking can be derived as
\begin{eqnarray}\label{5}
\begin{aligned}
\delta&=6K_{\mathrm{a}}(t)\!-\!2K_{\mathrm{m}}(t), \\ \Delta&=2[K_{\mathrm{a}}(t)\!+\!K_{\mathrm{m}}(t)\!-\!K_{\mathrm{t}}(t)].
\end{aligned}
\end{eqnarray}
With the help of condition~(\ref{5}), the system of equations~(\ref{2}) becomes
\begin{eqnarray}\label{6}
\begin{aligned}
i\dot{\psi}_{\mathrm{a}}\!&=\!-\Omega_{1}{\psi_{\mathrm{m}}\psi^{\ast}_{\mathrm{a}}}^{2}\!+\!\Omega_{2}\psi_{\mathrm{t}}{\psi^{*}_{\mathrm{m}}},\\ i\dot{\psi}_{\mathrm{m}}\!&=\!-\!i\gamma_\mathrm{m}\psi_\mathrm{m}\!-\Omega_{1}\psi_{\mathrm{a}}^{3}\!+\!\Omega_{2}\psi_{\mathrm{t}}{\psi^{\ast}_{\mathrm{a}}}, \\
i\dot{\psi}_{\mathrm{t}}\!&=\!-\!i\gamma_\mathrm{t}\psi_\mathrm{t}+\!\Omega_{2}\psi_{\mathrm{m}}\psi_{\mathrm{a}}.
\end{aligned}
\end{eqnarray}
Now two optical fields $\Omega_{1, 2}$ are resonant with the transitions $|\mathrm{a}\rangle \leftrightarrow |\mathrm{m}\rangle$ and
$|\mathrm{m}\rangle \leftrightarrow |\mathrm{t}\rangle$, respectively. The detailed derivation from Eq.~(\ref{2}) to~(\ref{6}) is presented in the~\hyperref[app1]{\textcolor{blue}{Appendix A}}.

We further note that, in our model~(\ref{6}), the couplings of both fields are nonlinear, which differs from Ref.~\cite{PhysRevLett.119.243902}, where only the coupling of the first field is nonlinear. In what follows, we demonstrate how the generalized STIREP adapts to the system to compensate for the nonlinearities induced by these couplings. Consistent with previous theoretical treatments, we ignore all decay rates in Eq.~(\ref{6}) to clarify the underlying physics. For the case with the absence of losses ($\gamma_{m, t}\!=\!0$), we have  $|\psi_\mathrm{a}|^2\!+\!3|\psi_\mathrm{m}|^2\!+\!4|\psi_\mathrm{t}|^2\!=\!1$. On this basis, we introduce two time-dependent angles $\theta(t)$ and $\phi(t)$ to parameterize the state vector:
\begin{eqnarray}\label{7}
\begin{aligned}
&\psi_{\mathrm{a}}\!=\!\cos\phi(t)\cos\theta(t),\\
&\psi_{\mathrm{m}}\!=\!-i\frac{\sin\phi(t)}{\sqrt{3}},\\
&\psi_{\mathrm{t}}\!=\!-\frac{\cos\phi(t)\sin\theta(t)}{2}.
\end{aligned}
\end{eqnarray}

Inserting this into Eqs.~(\ref{6}) we can get
\begin{equation}\label{8}
\begin{split}
\Omega_1(t)&\!=\!-\frac{\sqrt{3}}{12}\biggl(\frac{1\!+\!3\cos^{2}\theta(t)}{\cos^3\theta(t)\cos^2\phi(t)}\dot{\phi}(t)\\ &\qquad+\frac{10\tan\theta(t)\dot{\theta}(t)}{\sin2\phi(t)\cos\theta(t)}\biggr),\\
\Omega_2(t)&\!=\!\frac{\sqrt{3}}{2}\biggl(\frac{\dot{\theta}(t)}{\sin\phi(t)}\!-\!\frac{\dot{\phi}(t)\tan\theta(t)}{\cos\phi(t)}\biggl).
\end{split}
\end{equation}

To ensure the desired atom-to-tetratomic molecular conversion, we apply boundary conditions: $0\!\leftarrow\!\phi\!\rightarrow\!0$ and $0\!\leftarrow\!\theta\!\rightarrow\!\pi/2$,
where the arrows to the left and right signify the limits
when $t\!\rightarrow\!t_{i}$ and $t\!\rightarrow\!t_{f}$, respectively. The time-dependent angles $\theta(t)$ and $\phi(t)$ are chosen as
\begin{eqnarray}\label{9}
\begin{aligned}
\theta(t)&=\frac{\pi}{4}\eta\left[\mathrm{tanh}\left(\frac{t}{T}\right)\!+\!1\right], \\
\phi(t)&=\frac{4\epsilon\sqrt{\theta(t)\left[\frac{\pi}{2}\!-\!\theta(t)\right]}}{\pi},
\end{aligned}
\end{eqnarray}
where $\epsilon$ and $\eta$ are constants, $T$ serves as a time for normalization. $\epsilon\gtrsim0$ enables the regulation of the transient population in the intermediate state, while $\eta\lesssim1$ governs the final conversion efficiency. It should be noted that $\eta=1$ is prohibited in nonlinear cases, whereas no such restriction exists in linear cases. Once $\theta(t)$ and $\phi(t)$ are determined, the Rabi frequencies can be derived by substituting them into Eq.~(\ref{8}).
\subsection{\label{sec:level2.2}Result and Discussion}
\begin{figure}[t]
\centering{\includegraphics[width=8cm]{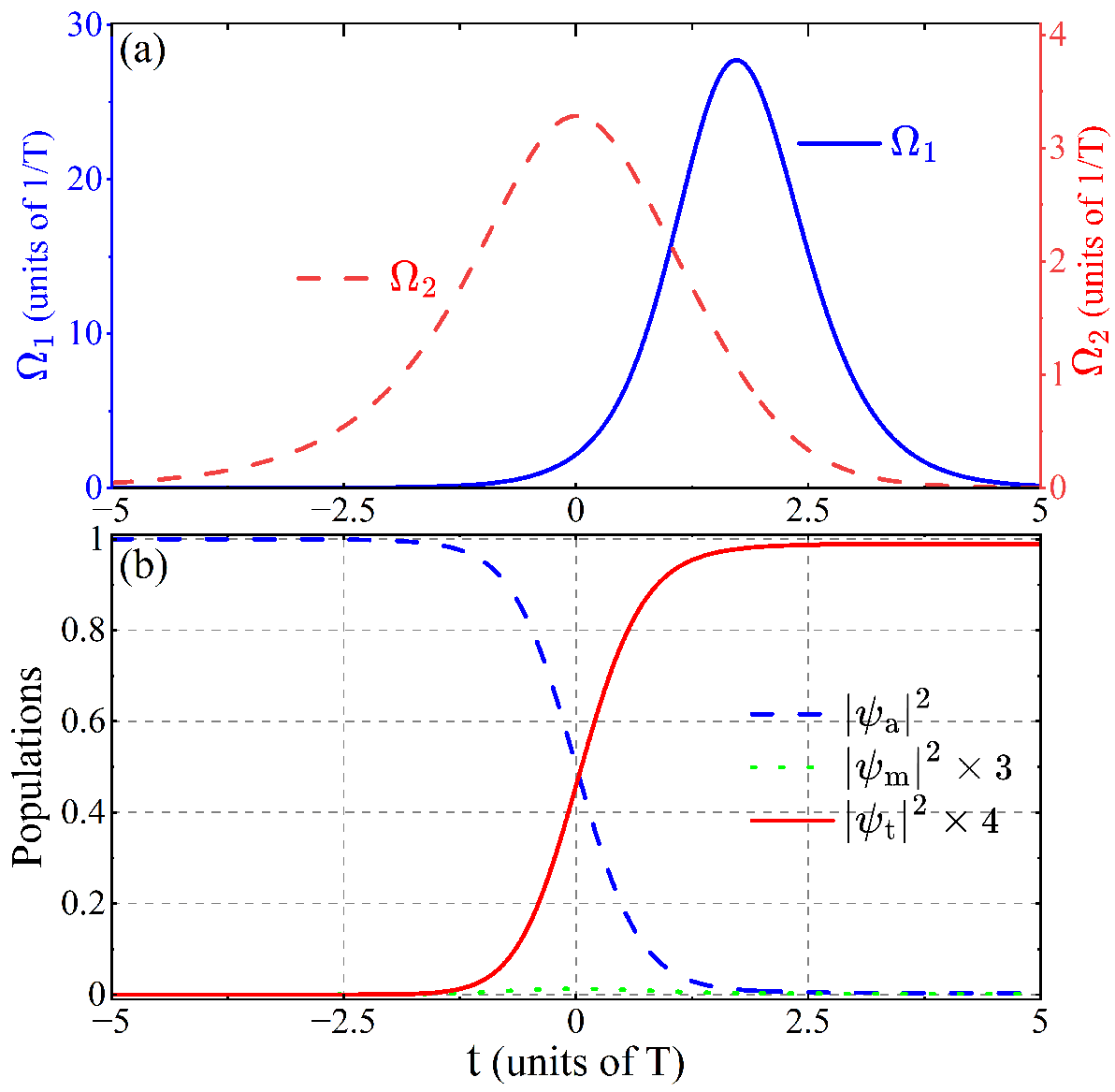}}
\caption{(Color online) Rabi frequencies (top panel) and populations (bottom panel) as a function of $t$. Adopted Parameters: $\eta=0.96, \epsilon=0.2, \gamma_{\mathrm{m,t}}=0$.}
\label{fig2}
\end{figure}
To demonstrate the validity of our procedure, we are going to carry out numerical calculations for population dynamics using Eqs.~(\ref{6}),~(\ref{8}) and~(\ref{9}), with the standard fourth-order Runge-Kutta method.

Figure.~\ref{fig2}(a) shows the time-dependent evolution of Rabi frequencies used for the conversion of atoms to tetratomic molecules. In Fig.~\ref{fig2}(b), we present the corresponding population dynamics. As anticipated, we achieved an almost perfect conversion efficiency, while maintaining a small transient population in the intermediate state during the interaction. Clearly, our method has the potential to maintain the phase-space density of ultracold mixtures, as a high conversion efficiency is essential for this maintenance. 

\begin{figure}[t]
\centering{\includegraphics[width=8.5cm]{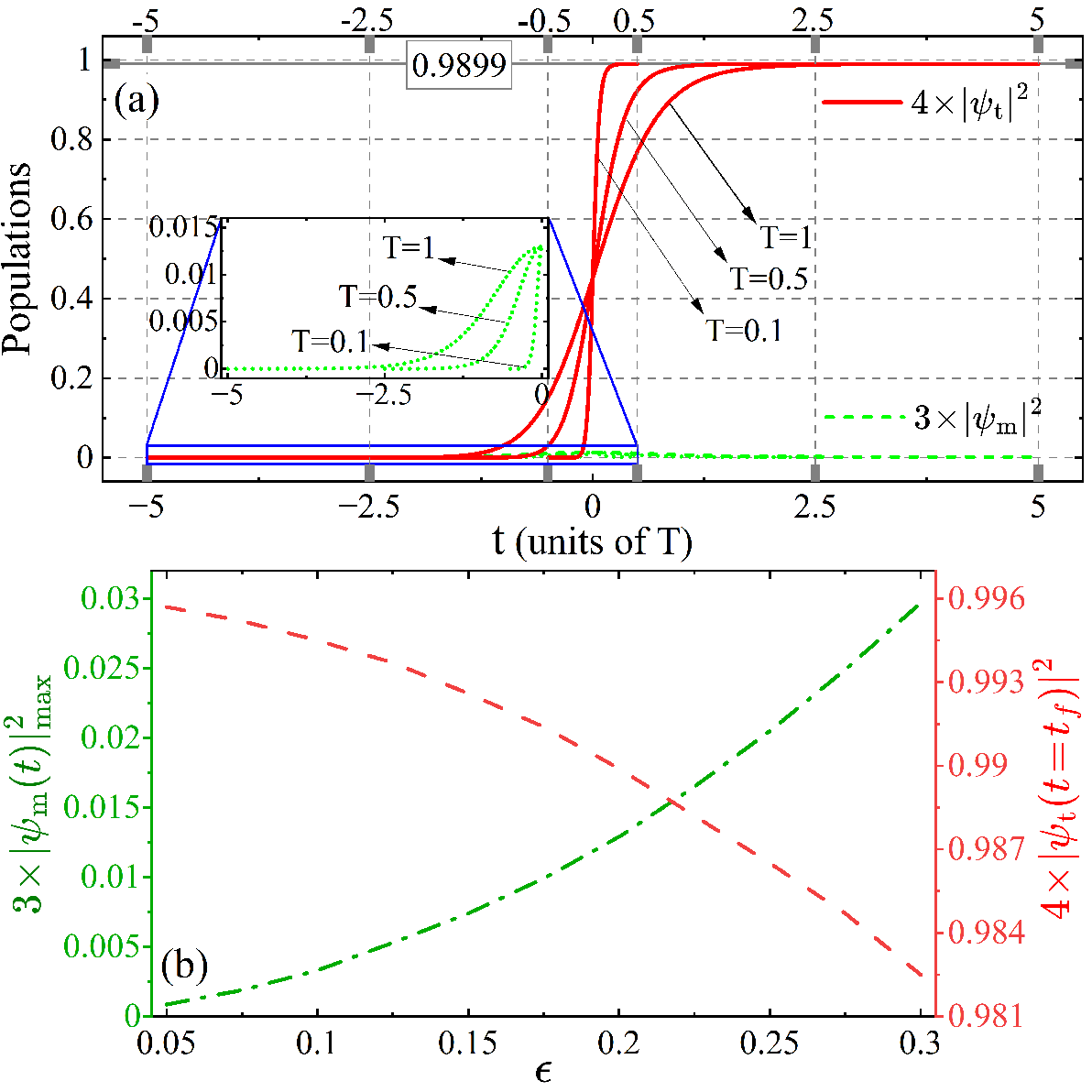}}
\caption{(Color online) The maximum transient population of intermediate triatomic molecules and the final population of tetratomic molecules as a function of parameter $\epsilon$. All other parameters are consistent with those in Fig.~\ref{fig2}.}
\label{fig3}
\end{figure}

In addition, we stress that our protocol allows for faster control over the conversion process at the same conversion efficiency. We demonstrate this fact by plotting a curve of the conversion efficiency of our protocol as a function of the evolution time $T$, as depicted by the solid red line in Fig.~\ref{fig3}(a). This intriguing phenomenon aligns with the core spirit of STA. Owing to the exact dynamics, we find that although the maximum transient population of the intermediate triatomic molecular state is independent of the value of $T$, the total transient population of this state over the entire process can be reduced by shortening the evolution time $T$. At the same time, given that $\epsilon$ is a key parameter governing the transient population of the intermediate triatomic state, we investigate the maximum transient population of the intermediate triatomic molecules and the final population of tetratomic molecules as a function of parameter $\epsilon$ in Fig.~\ref{fig3}(b). The results presented in this figure show that decreasing the value of parameter $\epsilon$ not only helps to improve the conversion efficiency $\small{4}|\psi_\mathrm{t}\small{(t\!=\!t_f)}|^2$, but also effectively suppresses the transient population of intermediate triatomic molecules. From the analysis of this figure, we can thus conclude that the decay of the intermediate triatomic state can be minimized via two ways, namely the reasonable selection of a small value of the parameter $\epsilon$, and the shortening of the evolution time.

To better illustrate the advantages of our protocol, we simulated its performance and compared it with that of the conventional STIRAP protocol. This study is feasible, because the transfer process we consider can be viewed as an abstract three-level model. In our simulations, the STIRAP process was implemented using a pair of Gaussian pulses,
\begin{eqnarray}\label{10}
\begin{aligned}
\Omega_{\mathrm{P, S}}(t)&=\Omega^{0}_{\mathrm{P, S}}\exp\left[-\frac{(t-t_{\mathrm{P, S}})^2}{T^2}\right],\\
\end{aligned}
\end{eqnarray}
where $T$, $t_{\mathrm{P,S}}$ and $\Omega^{0}_{\mathrm{P,S}}$ denote the width, peak times and peak strengths of the Gaussian pulses, respectively. The Stokes pulse $\Omega_{\mathrm{S}}$ is applied first to induce the transition $|\mathrm{m}\rangle \leftrightarrow |\mathrm{t}\rangle$ and is followed by the pump pulse $\Omega_{\mathrm{P}}$, which induces the transition $|\mathrm{a}\rangle \leftrightarrow |\mathrm{m}\rangle$.

\begin{figure}[b]
\centering
\begin{minipage}[b]{0.465\textwidth}
\includegraphics[width=\linewidth]{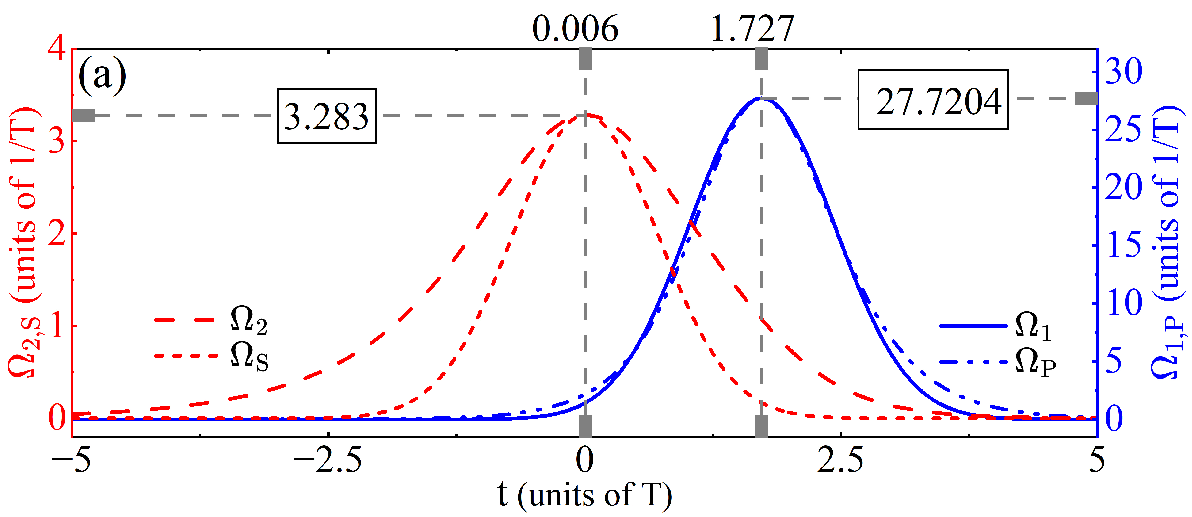}
\end{minipage}
\begin{minipage}[b]{0.465\textwidth}
\includegraphics[width=\linewidth]{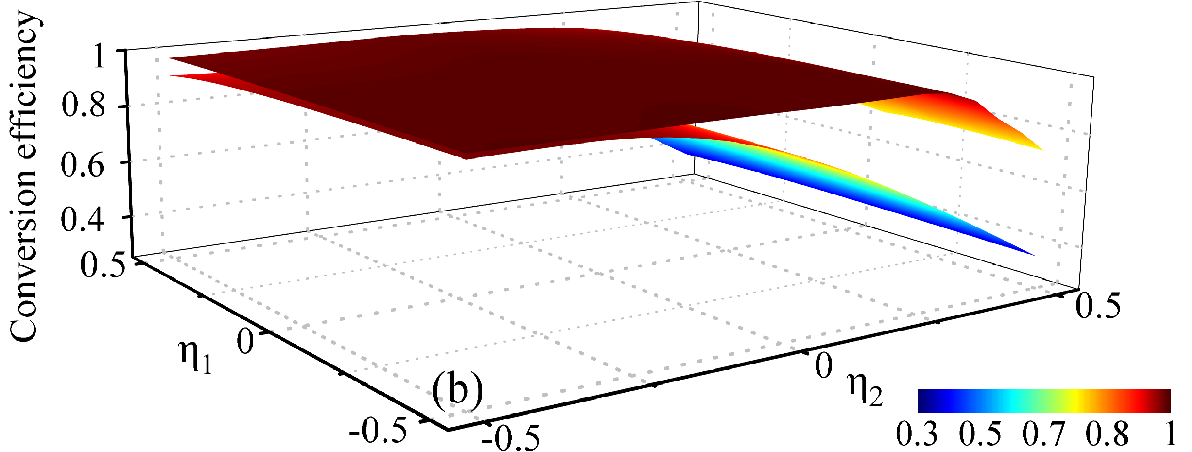}
\end{minipage}
\caption{(Color online) (a) Rabi frequencies as functions of $t$ under the generalized STIREP and the STIRAP. Adopted Parameters: $\eta=0.96, \epsilon=0.2$; (b) Conversion efficiency as a function of $\eta_1$ and $\eta_2$ under the (upper surface) generalized STIREP and (lower surface) STIRAP protocols.} 
  \label{fig4}
\end{figure}

\begin{figure}[t]
\centering{\includegraphics[width=7.5cm]{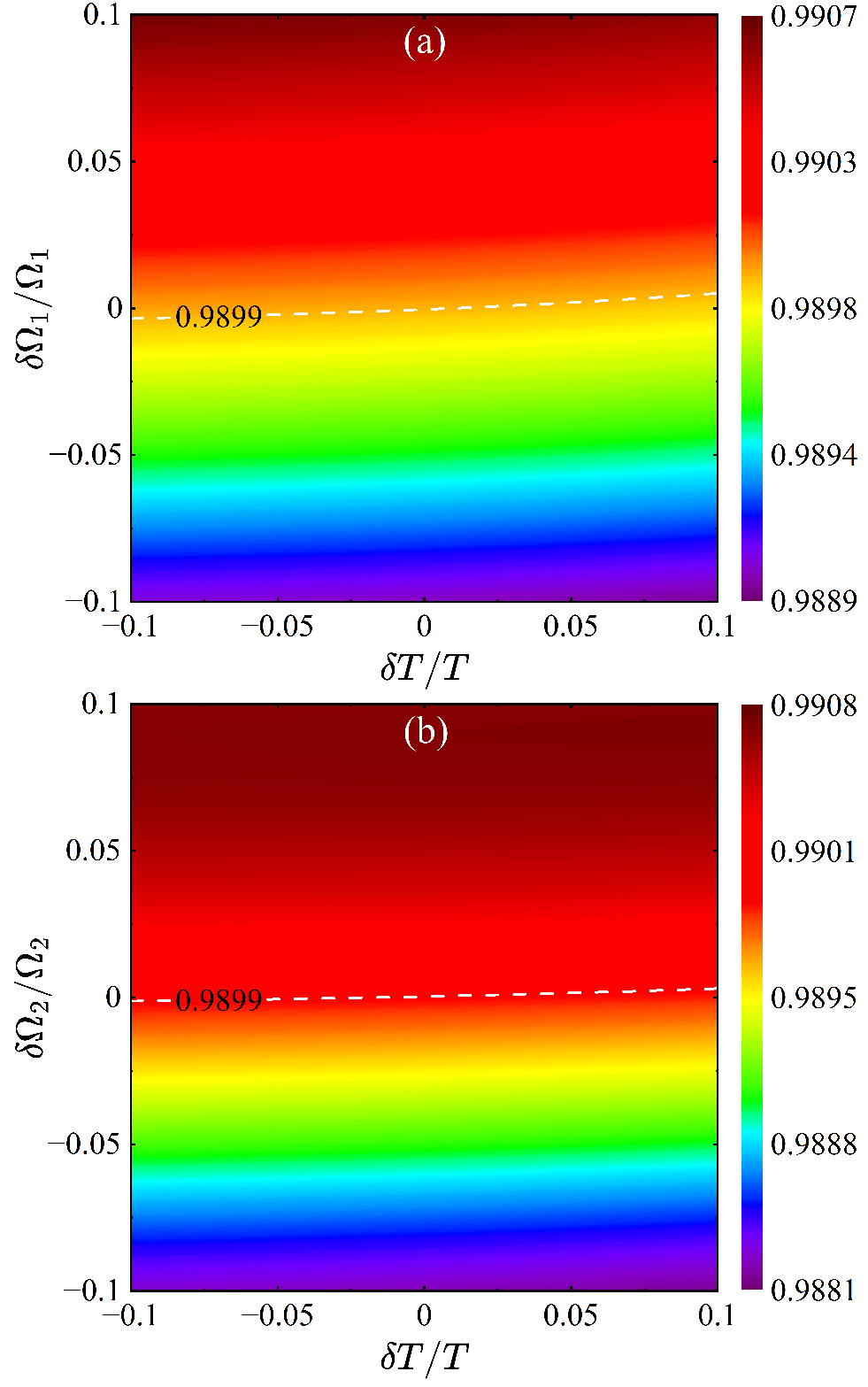}}
\caption{(Color online) (a) Contour plot showing conversion efficiency versus $\delta T$ and $\delta\Omega_1$, where $\Omega_2$ is fixed at its ideal value without any deviations; (b) Contour plot showing conversion efficiency versus $\delta T$ and $\delta\Omega_2$, where $\Omega_1$ is fixed at its ideal value. Note that parameter deviations are implemented on the basis of the original parameters in Fig.~\ref{fig2}.}
\label{fig5}
\end{figure}

To ensure a fair comparison, both schemes are carried out with the same evolution time, peak times, and peak strengths. According to Fig.~\ref{fig2}(a), where the curves are determined by Eq.~(\ref{8}) and Eq.~(\ref{9})  [also presented in Fig.~\ref{fig4}(a)], we accordingly set the amplitudes of the Gaussian envelopes to $\Omega^{0}_{\mathrm{P}}=27.7204/T $and $\Omega^{0}_{\mathrm{S}}=3.283/T$, and the peak times to $t_{\mathrm{P}}\!=\!1.727T $and $t_{\mathrm{S}}\!=\!0.006T$. Specifically, the amplitudes of the Gaussian envelopes are directly obtained from the vertical coordinates corresponding to the peaks of the curves, while the peak times are directly derived from the horizontal coordinates corresponding to these peaks.
Meanwhile, the robustness of both schemes against Rabi frequency errors will be assessed. To this end, we systematically modify the amplitudes of the original Rabi frequencies while preserving their waveform shapes, with specific adjustments defined as $\Omega_{\mathrm{1, P}}\rightarrow(1\!-\!\eta_1)\Omega_{\mathrm{1, P}}$ and $\Omega_{\mathrm{2, S}}\rightarrow(1\!-\!\eta_2)\Omega_{\mathrm{2, S}}$, respectively.

Figure.~\ref{fig4}(a) displays the time dependence of the Rabi frequencies corresponding to our protocol and STIRAP. It is obvious that the fields used in our protocol exhibit profiles with sufficient smoothness. We plot in Fig.~\ref{fig4}(b) the conversion efficiency as a function of $\eta_1$ and $\eta_2$. Here, the upper surface corresponds to our protocol, whereas the lower surface corresponds to STIRAP.
As it is observed, our protocol always performs better than STIRAP. 

Further studies can reveal more properties of our protocol. In Fig.~\ref{fig5}(a), by fixing $\Omega_2$ at its ideal value, we present a contour plot of transfer efficiency against the evolution time deviation $\delta T$ and the Rabi frequency deviation $\delta\Omega_1$. For any physical quantity $X$, its deviation is explicitly defined in~\hyperref[app2]{\textcolor{blue}{Appendix B}}. As a comparison, we study in Fig.~\ref{fig5}(b) transfer efficiency against $\delta T$ and $\delta\Omega_2$, with $\Omega_1$ fixed at its ideal value. It should be noted that the values marked by the white dashed lines in these figures are equal to the ideal value under the working condition of no deviations; this value correspond exactly to the end value of the red curve in Fig.~\ref{fig2}(b), and its specific value is $0.9899$. It is observed that even for deviations of $|\delta T/T|\!=\!|\delta\Omega_1/\Omega_1|\!=\!|\delta\Omega_2/\Omega_2|\!=\!10\%$, the conversion efficiencies remain remarkably close to the ideal value. The results demonstrate that our protocol exhibits high robustness with respect to imperfections in operation.

The above calculations demonstrate that the robustness, efficiency, and fast conversion of our protocol facilitate the preparation of ultracold tetratomic molecules.
Nevertheless, it should be stressed that the resulting molecules are usually energetically unstable~\cite{Kuznetsova_2009}, thereby making it highly necessary from a practical standpoint to rapidly transfer them to deeply-bound vibrational states~\cite{Danzl2010, PhysRevA.78.021402, j17b-x1x7}. To this end, we suggest a five-state C-STIREP technique,
whose details will be further analyzed and presented in the following section.

\section{\label{sec:level3}Ground-state molecular population transfer via C-STIREP}

\subsection{\label{sec:level3.1}Model and Method}
\begin{figure}[b]
\centering{\includegraphics[width=7.5cm]{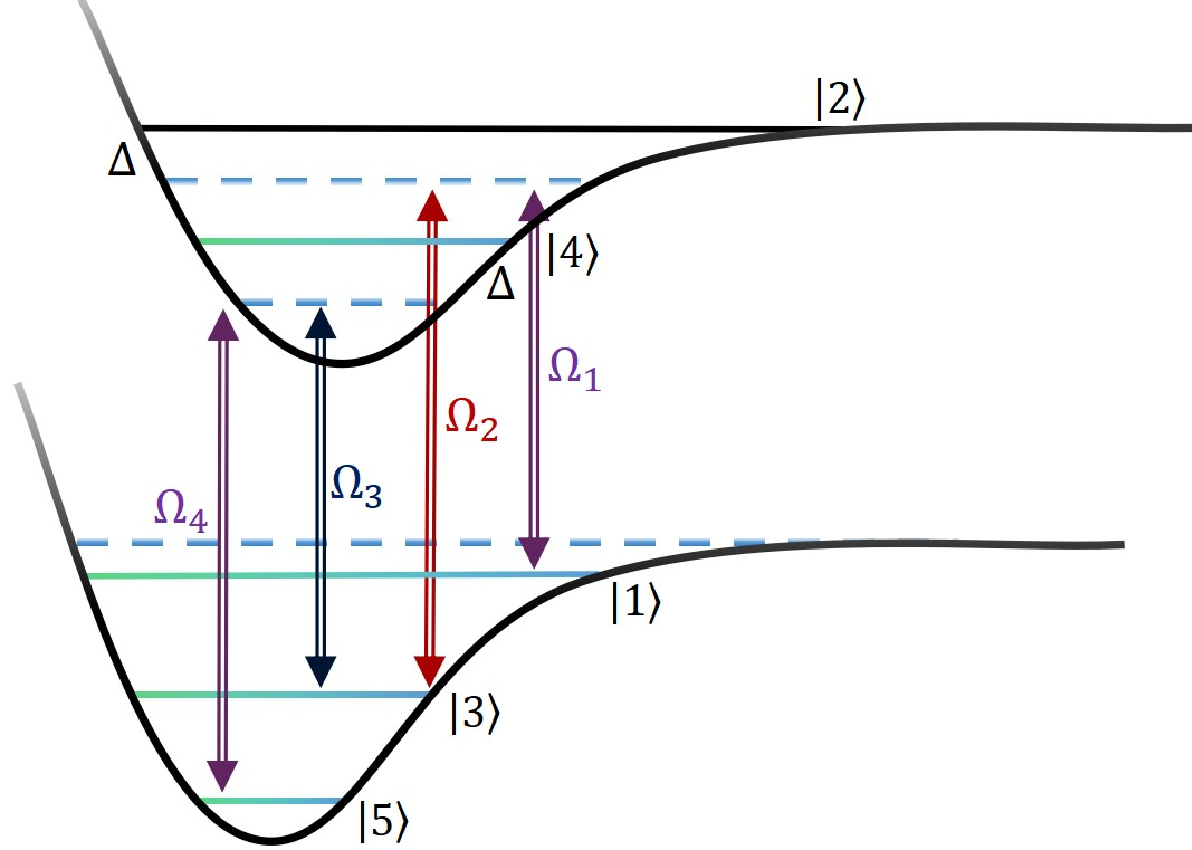}}
\caption{(Color online) A schematic of coherent population transfer from weakly- to deeply-bound molecules.}
\label{fig6}
\end{figure}

In this section, our goal is to generalize the STIREP into multi-state system, with the motivation to transfer the weakly-bound molecules into a sufficiently stable state.
Inspired by the double-STIRAP transfer scheme~\cite{Danzl2010, PhysRevA.78.021402}, we restrict ourselves to a five-level chainwise-coupled molecular system.
As presented in Fig.~\ref{fig6}, all five molecular states are coupled in a distorted M-type configuration,
in which four fields with Rabi frequencies $\Omega_i (i=1,2,3,4)$ are applied to move the molecules from a high vibrational level of the ground electronic state $|c\rangle$ (now relabeled as $|1\rangle$) to a sufficiently stable vibrational level of the ground electronic state $|5\rangle$. In the entire transfer process, three intermediate states are introduced to act as the ``bridge", where $|3\rangle$ is regarded as an intermediate vibrational level of ground electronic state, $|2\rangle$ and $|4\rangle$ represent two vibrational levels of the excited electronic state.
To render this problem tractable, we shall analyze this problem under the collisionless limit, as done in prior Refs.~\cite{PhysRevA.78.033617, PhysRevLett.98.050406, PhysRevA.81.013632, PhysRevA.92.012121}.
Accordingly, the Hamiltonian of the system can be described in the following form:
\begin{eqnarray}\label{11}
\begin{aligned}
\hat{\mathcal{H}}\!=\!\sum_{i=2}^{5}\Delta_{i}\hat{\psi}_{i}^{\dagger}\hat{\psi}_{i}
\!+\!\sum_{i=1}^{4}\left(\Omega_{i}\hat{\psi}_{i}^{\dagger}\hat{\psi}_{i+1}\!+\!\mathrm{H.c.}\right).
\end{aligned}
\end{eqnarray}
The Heisenberg motional equation can be written as
\begin{eqnarray}\label{12}
\begin{aligned}
i\dot{\psi}_{1}&\!=\!\Omega_{1}\psi_{2},\\
i\dot{\psi}_{2}&\!=\!\Omega_{1}\psi_{1}\!+\!\Delta\psi_{2}\!+\!\Omega_{2}\psi_{3},\\
i\dot{\psi}_{3}&\!=\!\Omega_{2}\psi_{2}\!+\!\Omega_{3}\psi_{4},\\
i\dot{\psi}_{4}&\!=\!\Omega_{3}\psi_{3}\!+\!\Delta\psi_{4}\!+\!\Omega_{4}\psi_{5},\\
i\dot{\psi}_{5}&\!=\!\Omega_{4}\psi_{4}.
\end{aligned}
\end{eqnarray}
For simplicity, we have assumed that $\Delta_{3}\!=\Delta_{5}\!=\!0$, and $\Delta_2\!=\!\Delta_4\!=\!\Delta$.
In what follows, we elaborate on how the C-STIREP technique operates within the above system.

First, we postulate that $\Delta$ is the largest evolution frequency~\cite{PhysRevX.9.021039}, $\Delta\!\gg\!\Omega_i (i\!=\!1,2,3,4)$.
In this limit, adiabatic elimination (AE) enables the derivation of an effective three-state model,
\begin{eqnarray}\label{13}
\begin{aligned}
i\dot{\psi}_{1}&=\varepsilon_{11}\psi_{1}\!+\!\omega_{13}\psi_{3},\\
i\dot{\psi}_{3}&=\varepsilon_{22}\psi_{3}\!+\!\omega_{13}\psi_{1}\!+\!\omega_{35}\psi_{5},\\
i\dot{\psi}_{5}&=\varepsilon_{33}\psi_{5}\!+\!\omega_{35}\psi_{3}.
\end{aligned}
\end{eqnarray}
In which, the effective Rabi frequencies are
$\omega_{13}=\!-\!\frac{\Omega_{1}\Omega_{2}}{\Delta}$ and $\omega_{35}=\!-\!\frac{\Omega_{3}\Omega_{4}}{\Delta}$;
the effective detunings are
$\varepsilon_{11}=\!-\!\frac{\Omega_{1}^2}{\Delta},
\varepsilon_{22}=\!-\!\frac{\Omega_{2}^2+\Omega_{3}^2}{\Delta}$ and
$\varepsilon_{33}=\!-\!\frac{\Omega_{4}^2}{\Delta}$, respectively.
To continue, let us set $\varepsilon_{11}=\varepsilon_{22}=\varepsilon_{33}=\varepsilon_0$,
from which the condition for generalized resonance locking follows,
\begin{eqnarray}\label{14}
\Omega_{1}(t)=\Omega_{4}(t)=\sqrt{\Omega_{2}^2(t)+\Omega_{3}^2(t)}.
\end{eqnarray}

By further performing the simple transformation $\psi_{j}\!=\!\psi^{'}_{j}e^{-i\varepsilon_0 t}(j\!=\!1,3,5\!)$, we obtain
\begin{eqnarray}\label{15}
\begin{aligned}
i\dot{\psi}^{'}_{1}&\!=\!\omega_{13}\psi^{'}_{3},\\
i\dot{\psi}^{'}_{3}&\!=\!\omega_{13}\psi^{'}_{1}\!+\!\omega_{35}\psi^{'}_{5},\\
i\dot{\psi}^{'}_{5}&\!=\!\omega_{35}\psi^{'}_{3},
\end{aligned}
\end{eqnarray}
in which
\begin{eqnarray}\label{16}
\begin{aligned}
\omega_{13}\!&=-\!\frac{\Omega_{2}(t)\sqrt{\Omega_{2}^2(t)\!+\!\Omega_{3}^2(t)}}{\Delta},\\
\omega_{35}\!&=-\!\frac{\Omega_{3}(t)\sqrt{\Omega_{3}^2(t)\!+\!\Omega_{2}^2(t)}}{\Delta}.
\end{aligned}
\end{eqnarray}

Having obtaining the equivalent model~(\ref{15}), we immediately parameterize the state vector as
\begin{eqnarray}\label{17}
\begin{aligned}
&\psi^{'}_{1}\!=\!\cos\phi(t)\cos\theta(t),\\
&\psi^{'}_{3}\!=\!-i\sin\phi(t),\\
&\psi^{'}_{5}\!=\!-\cos\phi(t)\sin\theta(t),
\end{aligned}
\end{eqnarray}
The Rabi frequencies can be constructed as
\begin{eqnarray}\label{18}
\begin{aligned}
\omega_{13}&=\dot{\theta}(t)\cot\phi(t)\sin\theta(t)\!+\!\dot{\phi}(t)\cos\theta(t), \\
\omega_{35}&=\dot{\theta}(t)\cot\phi(t)\cos\theta(t)\!-\!\dot{\phi}(t)\sin\theta(t).
\end{aligned}
\end{eqnarray}
In order to drive the state transfer of interest, we apply boundary conditions $0\!\leftarrow\!\phi\!\rightarrow\!0, 0\!\leftarrow\!\theta\!\rightarrow\!\pi/2$.

Up to now, we have obtained the STIREP for the equivalent system~(\ref{15}). In principle, by means of the two fields given by Eq.~(\ref{18}), one could leverage the advantages of STIREP for achieving the desired population transfer; however, it should be noted that such two-photon transitions are likely to be challenging to induce directly in practice.
Thus, the remaining task is to engineer physically feasible driving fields within the five-level framework.

Now we are ready to design the modified Rabi frequencies for the original five-state system.
Like Eq.~(\ref{16}), we impose the following form on the two middle Rabi frequencies:
\begin{eqnarray}\label{19}
\begin{aligned}
\omega_{13}&=-\frac{\tilde{\Omega}_{2}(t)\sqrt{\tilde{\Omega}_{2}^2(t)\!+\!\tilde{\Omega}_{3}^2(t)}}{\Delta},\\
\omega_{35}&=-\frac{\tilde{\Omega}_{3}(t)\sqrt{\tilde{\Omega}_{2}^2(t)\!+\!\tilde{\Omega}_{3}^2(t)}}{\Delta}.
\end{aligned}
\end{eqnarray}
Meanwhile, we invoke the resonance locking condition~(\ref{14}) to modify the remaining two Rabi frequencies:
\begin{eqnarray}\label{20}
\begin{aligned}
\tilde{\Omega}_{1}(t)=\tilde{\Omega}_{4}(t)=\sqrt{\tilde{\Omega}_{2}^2(t)+\tilde{\Omega}_{3}^2(t)}.
\end{aligned}
\end{eqnarray}

Solving Eqs.~(\ref{19}) and~(\ref{20}), we thus obtain
\begin{eqnarray}\label{21}
\begin{aligned}
\tilde{\Omega}_{2}(t)&\!=\!\left(\frac{\Delta^2\omega^4_{13}}{\omega^2_{13}\!+\!\omega^2_{35}}\right)^{\frac{1}{4}},\\
\tilde{\Omega}_{3}(t)&\!=\!\left(\frac{\Delta^2\omega^4_{35}}{\omega^2_{13}\!+\!\omega^2_{35}}\right)^{\frac{1}{4}},\\
\tilde{\Omega}_{1}(t)&\!=\!\tilde{\Omega}_{4}(t)\!=\!\left[\Delta^2(\omega^2_{13}\!+\!\omega^2_{35})\right]^{\frac{1}{4}}.
\end{aligned}
\end{eqnarray}
With the newly designed Rabi frequencies, we thereby develop a five-state C-STIREP technique.
It should be noted that, to ensure the validity of this method, the AE condition $\Delta\!\gg\!\tilde{\Omega}_i(t) (i\!=\!1,2,3,4)$ should still be satisfied~\cite{PhysRevA.94.063411}.

\subsection{\label{sec:level3.2}Result and Discussion}
Now we numerically investigate the coherent population dynamics of the C-STIREP method.
Unlike Eq.~(\ref{9}) above, we demonstrate that the enhanced performance of C-STIREP can also be achieved by using other pulse shapes. As an example, we consider
\begin{eqnarray}\label{22}
\begin{aligned}
\phi(t)&=\frac{\beta}{2}\left[1-\cos(\frac{2\pi t}{T})\right],\\
\theta(t)&=\frac{\pi t}{2T}-\frac{1}{3}\sin(\frac{2\pi t}{T})+\frac{1}{24}\sin(\frac{4\pi t}{T}).
\end{aligned}
\end{eqnarray}
Once $\phi(t)$ and $\theta(t)$ are determined, the effective Rabi frequencies can be found according to Eq.~(\ref{18}).

\begin{figure}[t]
\centering{\includegraphics[width=8cm]{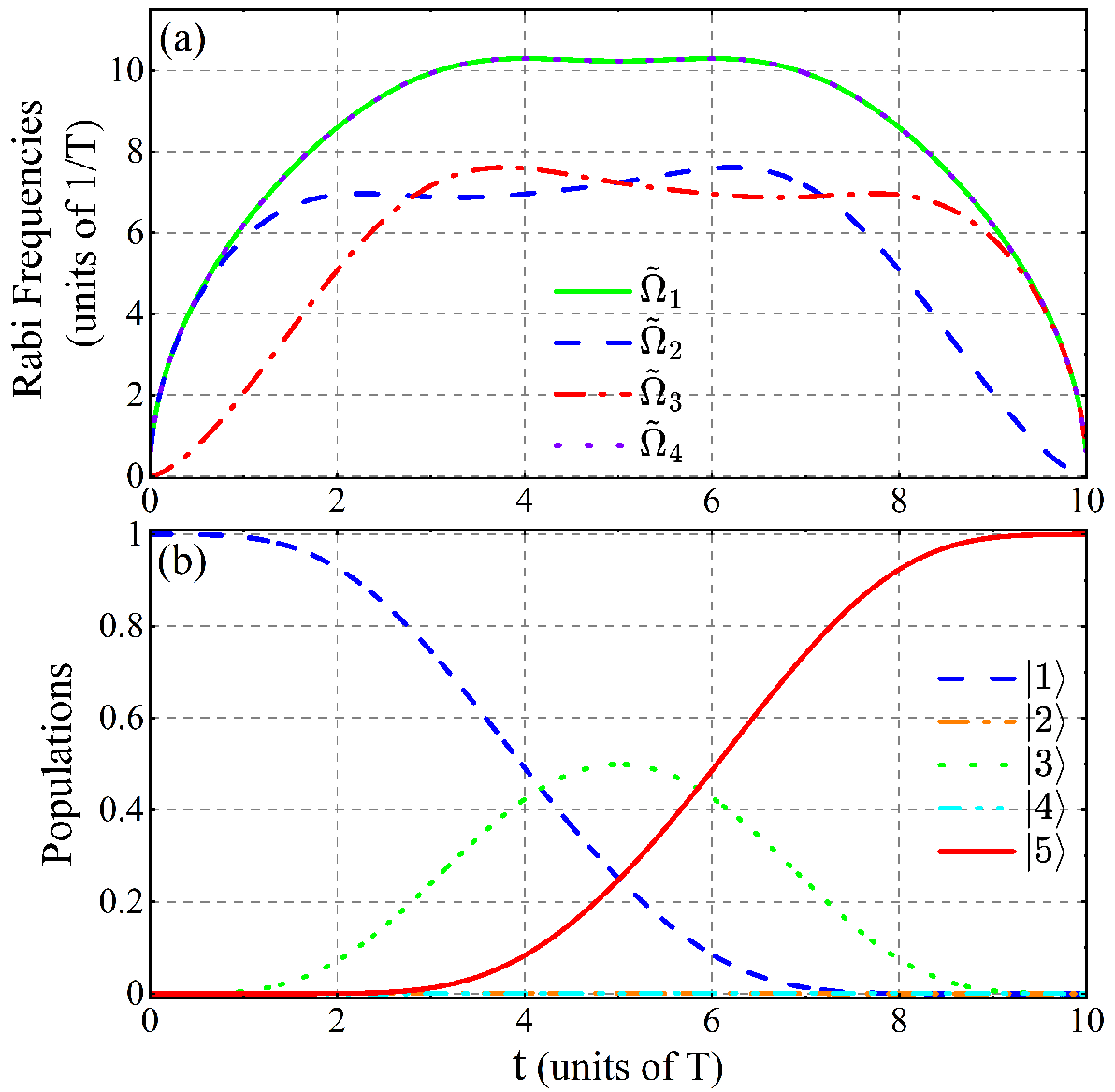}}
\caption{(Color online) (a) The sequence of four Rabi frequencies and (b) the population evolution of the five-state molecular system under the C-STIRAP. Adopted parameters: $\Delta\!=\!2500/T, \beta\!=\!0.25\pi$.}
\label{fig7}
\end{figure}
Figure.~\ref{fig7}(a) illustrates the Rabi frequencies required to achieve the desired population transfer. From this figure, it can be observed that the first and fourth Rabi frequencies not only exhibit the same temporal profile and equal magnitudes, but also straddle the middle two. They exhibit profiles with sufficient smoothness, which thus poses no actual implementation challenges~\cite{Du2016}.
The time evolution results from the initial state $|1\rangle$ to the target state $|5\rangle$ is displayed in Fig.~\ref{fig7}(b). It can be observed that a complete state transfer can be achieved by employing four modified fields. A high transfer efficiency is helpful to minimize the loss in phase-space density when an ensemble of weakly-bound molecules is transferred to the ground state. In addition, we note that, throughout the entire transfer process, the intermediate state $|3\rangle$ receives some amount of transient population, while the populations of the two excited states are negligible. This process does not rely on the existence of dark states; instead, we utilize the AE protocol to achieve the stable decoupling of two excited states from dynamics.
Thus, although these two states are prone to decaying out of the system, the transfer process remains scarcely swayed by them.

\begin{figure}[b]
\centering{\includegraphics[width=8cm]{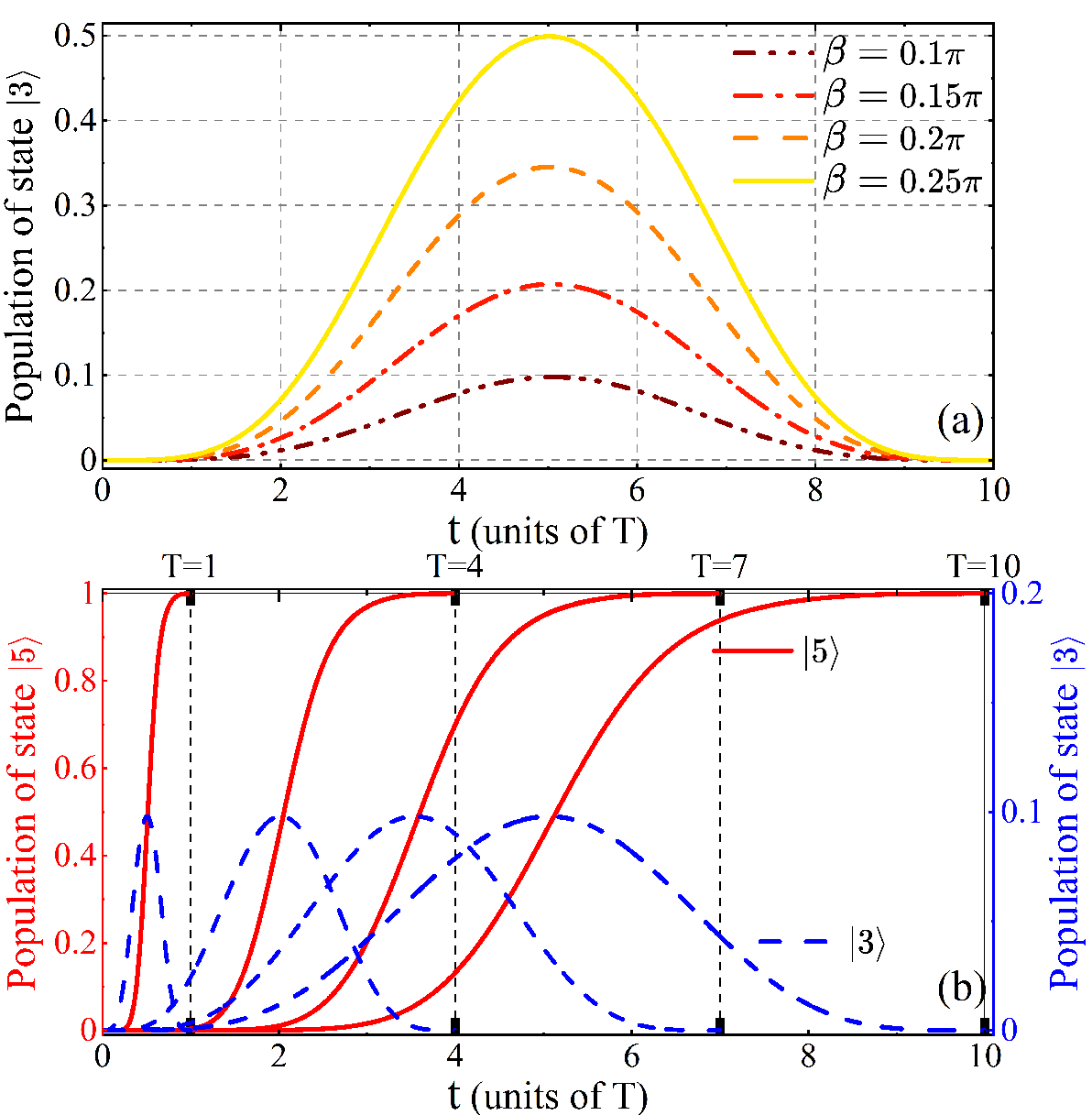}}
\caption{(Color online) (a) The transient population of state $|3\rangle$ as a function of $\beta$, where the other parameters are consistent with those in Fig.~\ref{fig7}; (b) The populations of state $|3\rangle$ and state $|5\rangle$ as a function of $t$ for different values of $T$, where $\beta$ is fixed at $0.1\pi$, and all other parameters are consistent with those in Fig.~\ref{fig7}.}
\label{fig8}
\end{figure}
Once a desired population transfer from $|1\rangle$ to $|5\rangle$ is achieved, we would like to impose further conditions of minimizing the amount of transient population that is in $|3\rangle$ throughout the process. Since the dynamics of the five-state system is equivalent to that of the reduced three-state one, the transient population of the intermediate state $|3\rangle$ can be accurately derived as $\sin^2\phi$. On this basis, we may attempt to reduce the value of $\phi$ to diminish the population of this state.
We plot in Fig.~\ref{fig8}(a) the transient population of state $|3\rangle$ as a function of $\beta$; please note that all other parameters are the same as those in Fig.~\ref{fig7}. Clearly, the population of the intermediate state $|3\rangle$ witnesses a substantial decrease as the value of $\beta$ is gradually reduced. However, the population of this state never equals $0$ since the limit of $\beta\rightarrow0$ cannot be achieved in practical scenarios [see Eqs.~(\ref{21}) and~(\ref{22})].
From the preceding analysis, it can be concluded that the five-state transfer scheme offers an additional benefit over the three-state counterpart: the former only requires us to control the transient population of the intermediate state $|3\rangle$, while the latter necessitates the control of the short-lived excited state. Given that the intermediate state $|3\rangle$ typically exhibits a sufficiently long lifetime, even a small amount of transient population within this state exerts a negligible influence on the final transfer efficiency; in contrast, the transfer efficiency of the three-state scheme is largely constrained by the characteristics of its intermediate short-lived excited state~\cite{PhysRevA.96.013406, Li:17}.

\begin{figure}[t]
\centering{\includegraphics[width=7.5cm]{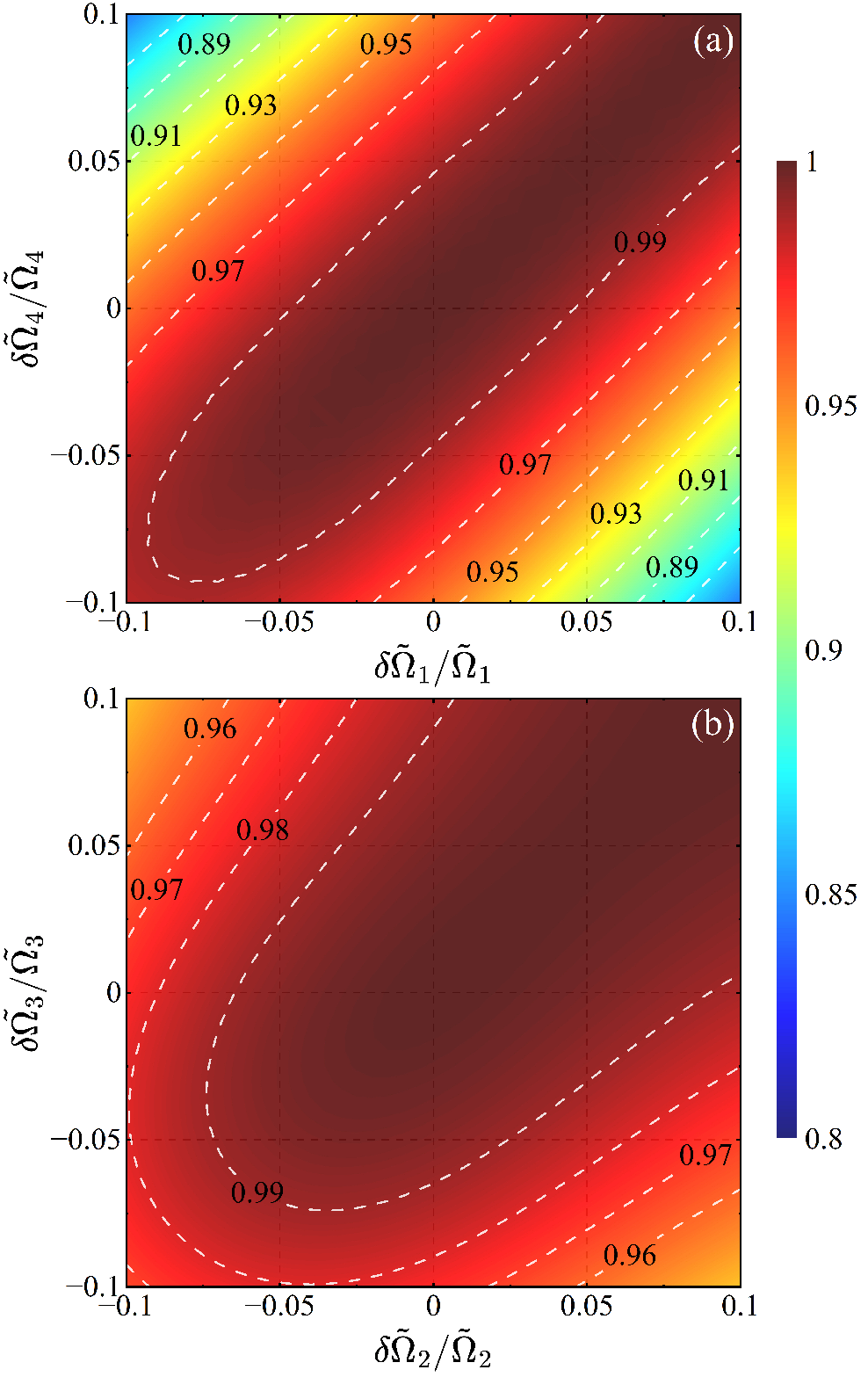}}
\caption{(Color online) (a) Contour plot showing transfer efficiency versus $\delta\tilde\Omega_1$ and $\delta\tilde\Omega_4$, where $\tilde{\Omega}_2$ and $\tilde{\Omega}_3$ are set to their ideal values; (b) Contour plot showing transfer efficiency versus $\delta\tilde\Omega_2$ and $\delta\tilde\Omega_3$, where $\tilde{\Omega}_1$ and $\tilde{\Omega}_4$ passively vary according to condition~(\ref{20}). The parameter deviations are implemented on the basis of the adopted parameters in Fig.~\ref{fig8}(b).}
\label{fig9}
\end{figure}
Furthermore, we point out that the transient population of the intermediate state $|3\rangle$ can be further reduced by leveraging the speed advantage of C-STIREP. To verify this finding, we plot in Fig.~\ref{fig8}(b) the variation of populations at time $t\!=\!T$ of states $|3\rangle$ and $|5\rangle$ with the evolution time $T$. As can be seen from this figure, the proposed method is capable of significantly shortening the population transfer time without deteriorating the transfer efficiency. This interesting feature mitigates the interaction duration between the system and its surrounding environment, thereby enhancing the transfer efficiency for practical applications.

Finally, we have performed numerical simulations to verify the necessity of Condition~(\ref{20}).
Figure.~\ref{fig9}(a) shows a contour plot of transfer efficiency against the Rabi frequency deviations $\delta\tilde\Omega_1$ and $\delta\tilde\Omega_4$ when $\tilde\Omega_2$ and $\tilde\Omega_3$ are fixed at their ideal values. The transfer efficiency is defined as
\begin{eqnarray}\label{23}
\begin{aligned}
\digamma\!=\!\frac{P_{5,t\rightarrow\infty}}{P_{1,t\rightarrow0}},
\end{aligned}
\end{eqnarray}
where $P_{i,t}$ denotes the population of state $|i\rangle$ at time $t$. Since the probability normalization and the conservation of the total particle number hold, $\digamma$ corresponds, in essence, to the population of state $|5\rangle$ at $t\!=\!T$. 
According to the results shown in this figure, we find that any deviation from the ideal condition will result in a significant drop in efficiency. In particular, when the deviation is $\pm 10\%$, the transfer efficiency will suffer a loss of around $10\%$.
Except for this, we can observe that the deviations $\tilde\Omega_1$ and $\tilde\Omega_4$ exert symmetric effects on the transfer efficiency.
This confirms that this condition is an indispensable prerequisite for obtaining high population transfer. 

Figure.~\ref{fig9}(b) presents a contour plot of the transfer efficiency as a function of
$\delta\tilde\Omega_2$ and $\delta\tilde\Omega_3$, with Condition~(\ref{20}) maintained; in this case,
$\tilde\Omega_1$ and $\tilde\Omega_4$ will passively vary with the other two. It is clear that the transfer efficiency can keep at a high level, when the deviation of the Rabi frequency deviations is within $|\delta\tilde\Omega_2/\tilde\Omega_2|\!=\!|\delta\tilde\Omega_3/\tilde\Omega_3|\!=\!10\%$. This result further demonstrates the necessity of Condition~(\ref{20}).

\section{\label{sec:level4}Conclusions and Outlooks}
To conclude, we present a theoretical formalism for the formation of stable ultracold tetratomic molecules.
The process begins with a mixture of ultracold atoms: first, generalized nonlinear STIREP is employed to coherently assemble atoms into tetratomic molecules; subsequently, to ensure the resulting molecules exhibit sufficient stability, we propose that the C-STIREP technique be used to transfer them to the stable ground state.
Numerical analyses show that the proposed two-step strategy enables the production of stable ultracold molecules with nearly ideal efficiency, while featuring excellent accelerability and high robustness. Although this is a theoretical study, we hope that with the advancement of experimental techniques, particularly in the formation of long-lived triatomic and tetratomic molecules, this two-step strategy will be translated into practical applications in the years to come~\cite{PhysRevLett.107.073201, doi:10.1126/science.aaa5601, doi:10.1126/science.ade6307, j17b-x1x7}.

The main advantages of our work are: (\romannum{1}) It paves a potential pathway for forming stable ultracold tetratomic molecules, and in principle, has the potential to be extended to preparing larger polyatomic molecular complexes~\cite{PhysRevA.87.043631}; (\romannum{2}) Unlike traditional methods, the adiabatic condition is released in our approach, thus enabling fast speed and high efficiency. Additionally, the transient populations of short-lived states can be effectively suppressed without relying on the dark state; (\romannum{3}) Although the protocol follows the spirit of STA, it does not introduce extra couplings, and the applied pulses pose no actual implementation difficulties; (\romannum{4}) Numerous exact solutions support the achievement of predefined goals, and the above-discussed examples~(\ref{9}) and~(\ref{22}) are not exhaustive.

Finally, we point out that the tetratomic molecule preparation process suggested in this study aligns closely with the standard procedure for diatomic molecule, with both following the similar route of ``first creating unstable molecular samples by assembling pre-cooled atom, and then converting them to sufficiently stable ground state". Inspired by the studies in Refs.~\cite{PhysRevA.82.011609, PhysRevA.100.033421}, we will, in
the future, commit to exploring a coherent strategy that starts directly from ultracold atoms and enables the robust, fast and efficient production of stable UPMs.


\section*{Acknowledgments}
We would like to thank the anonymous referees for constructive comments that are helpful for improving the quality of the work.
This work was supported by the National Natural Science Foundation of China (Grants No.~12475026, No.~12075193, and No.~12365004) and the Doctoral Research Startup Fund of Northwest Normal University (Grant No.~6014/202503101301).

\appendix
\section*{Appendix A}\label{app1}
In the Appendix, we will present the detailed derivation process from Eq.~(\ref{2}) to~(\ref{6}).
According to Eq.~(\ref{3}), we have
\begin{eqnarray}\label{A1}
\begin{aligned}
\dot{\psi}_{\mathrm{a}}&\!=\!\dot{\psi}_{\mathrm{a}}e^{i\alpha(t)}\!+\!i\dot{\alpha}(t)\psi_{\mathrm{a}}e^{i\alpha(t)}, \\
\dot{\psi}_{\mathrm{m}}&\!=\!\dot{\psi}_{\mathrm{m}}e^{3i\alpha(t)}\!+\!3i\dot{\alpha}(t)\psi_{\mathrm{m}}e^{3i\alpha(t)}, \\ \dot{\psi}_{\mathrm{t}}&\!=\!\dot{\psi}_{\mathrm{t}}e^{4i\alpha(t)}\!+\!4i\dot{\alpha}(t)\psi_{\mathrm{t}}e^{4i\alpha(t)}.
\end{aligned}
\end{eqnarray}
Substituting the above results into Eq.~(\ref{2}), we obtain
\begin{widetext}
\begin{eqnarray}\label{A2}
\begin{aligned}
i\left[\dot{\psi}_{\mathrm{a}}e^{i\alpha\left(t\right)}\!+\!i\dot{\alpha}\left(t\right)\psi_{\mathrm{a}}e^{i\alpha\left(t\right)}\right]\!&=\!-2K_\mathrm{a}\left(t\right)\psi_\mathrm{a}e^{i\alpha\left(t\right)}\!-\!3\Omega_{1}{\psi_\mathrm{m}e^{3i\alpha\left(t\right)}\psi^{\ast}_\mathrm{a}}^{2}e^{-2i\alpha\left(t\right)}\!+\!\Omega_{2}\psi_\mathrm{t}e^{4i\alpha\left(t\right)}{\psi^{*}_\mathrm{m}}e^{-3i\alpha\left(t\right)},\\
i\left[\dot{\psi}_{\mathrm{m}}e^{3i\alpha\left(t\right)}\!+\!3i\dot{\alpha}\left(t\right)\psi_{\mathrm{m}}e^{3i\alpha\left(t\right)}\right]&=\!-2K_\mathrm{m}\left(t\right)\psi_\mathrm{m}e^{3i\alpha\left(t\right)}\!-\!\left(\delta\!+\!i\gamma_\mathrm{m}\right)\psi_\mathrm{m}e^{3i\alpha\left(t\right)}\!-\!\Omega_{1}\psi_\mathrm{a}^{3}e^{3i\alpha\left(t\right)}\!+\!\Omega_{2}\psi_\mathrm{t}e^{4i\alpha\left(t\right)}{\psi^{\ast}_\mathrm{a}e^{-i\alpha\left(t\right)}},\\
i\left[\dot{\psi}_{\mathrm{t}}e^{4i\alpha\left(t\right)}\!+\!4i\dot{\alpha}\left(t\right)\psi_{\mathrm{t}}e^{4i\alpha\left(t\right)}\right]\!&=\!-2K_\mathrm{t}\left(t\right)\psi_\mathrm{t}e^{4i\alpha\left(t\right)}\!-\!\left(\Delta\!+\!\delta\!+\!i\gamma_\mathrm{t}\right)\psi_\mathrm{t}e^{4i\alpha\left(t\right)}\!+\!\Omega_{2}\psi_\mathrm{m}e^{3i\alpha\left(t\right)}\psi_\mathrm{a}e^{i\alpha\left(t\right)}.
\end{aligned}
\end{eqnarray}
\end{widetext}
It is evident that all terms on both sides of the above equation share a common factor of $e^{i\alpha(t)}$. By canceling out this common factor, the preceding equation can be simplified to:
\begin{widetext}
\begin{eqnarray}\label{A3}
\begin{aligned}
i\dot{\psi}_{\mathrm{a}}\!-\!\dot{\alpha}(t)\psi_{\mathrm{a}}\!&=\!-2K_{a}(t)\psi_\mathrm{a}\!-\!3\Omega_{1}{\psi_\mathrm{m}\psi^{\ast}_\mathrm{a}}^{2}\!+\!\Omega_{2}\psi_\mathrm{t}{\psi^{*}_\mathrm{m}},\\
i\dot{\psi}_{\mathrm{m}}\!-\!3\dot{\alpha}(t)\psi_{\mathrm{m}}\!&=\!-2K_\mathrm{m}(t)\psi_\mathrm{m}\!-\!(\delta\!+\!i\gamma_\mathrm{m})\psi_\mathrm{m}\!-\!\Omega_{1}\psi_\mathrm{a}^{3}\!+\!\Omega_{2}\psi_\mathrm{t}{\psi^{\ast}_\mathrm{a}},\\
i\dot{\psi}_{\mathrm{t}}\!-\!4\dot{\alpha}(t)\psi_{\mathrm{t}}\!&=\!-2K_\mathrm{t}(t)\psi_\mathrm{t}\!-\!(\Delta\!+\!\delta\!+\!i\gamma_\mathrm{t})\psi_\mathrm{t}\!+\!\Omega_{2}\psi_\mathrm{m}\psi_\mathrm{a}.
\end{aligned}
\end{eqnarray}
\end{widetext}
After moving all the like terms to one side, we can arrive at Eq.~(\ref{4}) presented in Section.~\ref{2.1}. Finally, by setting $\dot{\alpha}=2K_{a}(t)$ and performing some straightforward calculations, we end up with the condition for resonance locking~(\ref{5}); once this condition is satisfied, the system can be simplified to Eq.~(\ref{6}).

\appendix
\section*{Appendix B}\label{app2}
For any physical quantity $X$, its deviation is defined as
\begin{eqnarray}\label{24}
\begin{aligned}
\delta X=X\times\delta X/X,
\end{aligned}
\end{eqnarray}
where $X$ denotes its ideal value, and $\delta X/X$ represents the dimensionless relative deviation (ranging from $-0.1$ to $0.1$, i.e., $\pm10\%$). After incorporating the deviation, the realistic physical quantity used for calculation is given by $X\!+\!\delta X$.

\bibliography{references}
\end{document}